\documentclass[aps,showpacs,preprintnumbers,amsmath,amssymb]{revtex4}

  \usepackage{graphicx}

  \newcommand{\A}{{\cal A}}
  \newcommand{\D}{{\cal D}}

  \newcommand{\pplus}{p_{+}}
  \newcommand{\pmin}{p_{-}}
  \newcommand{\ppm}{p_\pm}

  \newcommand{\ie}{\emph{i.e.}}
  
  \newcommand{\cf}{\emph{cf.~}}

\begin{document}

\title{Exact solutions of $f(R)$ gravity coupled to nonlinear electrodynamics}

\author{Lukas Hollenstein}
\email{lukas.hollenstein@port.ac.uk}
\affiliation{Institute of Cosmology \& Gravitation,
             University of Portsmouth, Portsmouth PO1 2EG, UK}

\author{Francisco S. N. Lobo}%
\email{francisco.lobo@port.ac.uk} \affiliation{Institute of
Cosmology \& Gravitation, University of Portsmouth, Portsmouth PO1
2EG, UK}
%
%\affiliation{Centro de Astronomia e Astrof\'{\i}sica da
%Universidade de Lisboa, Campo Grande, Ed. C8 1749-016 Lisboa,
%Portugal}

\date{\today}

\begin{abstract}

In this work, exact solutions of static and spherically symmetric
space-times are analyzed in $f(R)$ modified theories of gravity
coupled to nonlinear electrodynamics. Firstly, we restrict the
metric fields to one degree of freedom, considering the specific
case of $g_{tt}\,g_{rr}=-1$. Using the dual $P$ formalism of
nonlinear electrodynamics an exact general solution is deduced in
terms of the structural function $H_P$. In particular, specific
exact solutions to the gravitational field equations are found,
confirming previous results and new pure electric field solutions
are found. Secondly, motivated by the existence of regular
electric fields at the center, and allowing for the case of
$g_{tt}\,g_{rr}\neq -1$, new specific solutions are found.
Finally, we outline alternative approaches by considering the
specific case of constant curvature, followed by the analysis of a
specific form for $f(R)$.

\end{abstract}

\maketitle

%-----------------------------------------------------------------
\section{Introduction}
%-----------------------------------------------------------------

A central theme in cosmology is the perplexing fact that the
Universe is undergoing an accelerated expansion \cite{expansion}.
Several candidates, responsible for this expansion, have been
proposed in the literature, in particular, dark energy models and
modified gravity. Amongst the modified theories of gravity, models
generalizing the Einstein-Hilbert action have been proposed, where
a nonlinear function of the curvature scalar, $f(R)$, is
introduced in the action. These modified theories of gravity seem
to provide a natural gravitational alternative to dark energy,
and in addition to allow for a unification of the early-time
inflation \cite{Starobinsky:1980te} and late-time cosmic speed-up
\cite{Carroll:2003wy,Fay:2007gg}. These models seem to explain the
four cosmological phases \cite{Nojiri:2006be}. They are also very
useful in high energy physics, in explaining the hierarchy problem
and the unification of GUTs with gravity \cite{Cognola:2006eg}.
The possibility that the galactic dynamics of massive test
particles may be understood without the need for dark matter was
also considered in the framework of $f(R)$ gravity models
\cite{darkmatter}. One may also generalize the action by
considering an explicit coupling between an arbitrary function of
the scalar curvature, $R$, and the Lagrangian density of matter
\cite{Nojiri:2004bi}. Note that these couplings imply the
violation of the equivalence principle \cite{Olmo:2006zu}, which
is highly constrained by solar system tests.

A fundamental issue extensively addressed in the literature is the
viability of the proposed $f(R)$ models
\cite{viablemodels,Hu:2007nk,Sokolowski:2007pk}. In this context,
it has been argued that most $f(R)$ models proposed so far in the
metric formalism violate weak field solar system constraints
\cite{solartests}, although viable models do exist
\cite{Hu:2007nk,solartests2,Sawicki:2007tf,Amendola:2007nt}. The
issue of stability \cite{Faraoni:2006sy} also plays an important
role for the viability of cosmological solutions
\cite{Nojiri:2006ri,Sokolowski:2007pk,Amendola:2007nt,Boehmer:2007tr,
deSitter}. In the context of cosmological structure formation
observations \cite{structureform}, it has been argued that the
inclusion of inhomogeneities is necessary to distinguish between
dark energy models and modified theories of gravity, and
therefore, the evolution of density perturbations and the study of
perturbation theory in $f(R)$ gravity is of considerable
importance \cite{Boehmer:2007tr,Tsuj,Uddin:2007gj,Bazeia:2007jj}.

A great deal of attention has also been paid to the issue of
static and spherically symmetric solutions of the gravitational
field equations in $f(R)$ gravity
\cite{Clifton:2005aj,SSSsol,Multamaki:2006zb}. Solutions in the
presence of a perfect fluid were also analyzed
\cite{Multamaki:2006ym}, where it was shown that the pressure and
energy density profiles do not uniquely determine $f(R)$. In
addition to this, it was found that matching the exterior
Schwarzschild-de Sitter metric to the interior metric leads to
additional constraints that severely limit the allowed fluid
configurations. An interesting approach in searching for exact
spherically symmetric solutions in $f(R)$ theories of gravity was
explored in \cite{Capozziello:2007wc}, via the Noether Symmetry
Approach, and a general analytic procedure was developed to deal
with the Newtonian limit of $f(R)$ gravity in
\cite{Capozziello:2007ms}. Analytical and numerical solutions of
the gravitational field equations for stellar configurations in
$f(R)$ gravity theories were also presented
\cite{Kainulainen:2007bt,Henttunen:2007bz,Multamaki:2007jk}, and
the generalized Tolman-Oppenheimer-Volkov equations for these
theories were derived \cite{Kainulainen:2007bt}.

In the context of $f(R)$ modified theories of gravity, it was
recently shown that power-law inflation and late-time cosmic
accelerated expansion can be explained by a modified
$f(R)$-Maxwell theory \cite{Bamba:2008ja}, due to breaking
the conformal invariance of the electromagnetic field through a
non-minimal gravitational coupling. It is interesting to note that
such a coupling may generate large-scale magnetic fields.
Motivated by these ideas, we consider in this work $f(R)$ gravity
coupled to nonlinear electrodynamics, and endeavor to search for
exact solutions in a static and spherically symmetric set-up. In
contrast to a non-minimal gravitational coupling, here conformal
invariance is not broken.

In the context of nonlinear electrodynamics, a specific model was
proposed by Born and Infeld in 1934 \cite{BI} founded on a
\emph{principle of finiteness}, namely, that a satisfactory theory
should avoid physical quantities to become infinite. The
Born-Infeld model was inspired mainly to remedy the fact that the
standard picture of a point particle possesses an infinite
self-energy, and consisted on placing an upper limit on the
electric field strength and considering a finite electron radius.
Later, Pleba\'{n}ski explored and presented other examples of
nonlinear electrodynamic Lagrangians \cite{Pleb}, and showed that
the Born-Infeld theory satisfies physically acceptable
requirements. Furthermore, nonlinear electrodynamics have recently
been revived, mainly because these theories appear as effective
theories at different levels of string/M-theory, in particular, in
D$p-$branes and supersymmetric extensions, and non-Abelian
generalizations (see Ref.~\cite{Witten} for a review).

Much interest in nonlinear electrodynamic theories has also been
aroused in applications to cosmological models \cite{cosmoNLED},
in particular, in explaining the inflationary epoch and the
late-time accelerated expansion of the universe \cite{Novello}. It
is interesting to note that the first {\it exact} regular black
hole solution in general relativity was found within nonlinear
electrodynamics \cite{Garcia,Garcia2}, where the source is a
nonlinear electrodynamic field satisfying the weak energy
condition, and recovering the Maxwell theory in the weak field
limit. In fact, general relativistic static and spherically
symmetric space-times coupled to nonlinear electrodynamics have
been extensively analyzed in the literature: regular magnetic
black holes and monopoles \cite{Bronnikov1}; regular electrically
charged structures, possessing a regular de Sitter center
\cite{Dymnikova}; traversable wormholes \cite{Arellano1} and
gravastar solutions \cite{Arellano3}.

Thus, as mentioned above, motivated by recent work on a
non-minimal Maxwell-$f(R)$ gravity model \cite{Bamba:2008ja}, in
this paper $f(R)$ modified theories of gravity coupled to
nonlinear electrodynamics are explored, in the context of static
and spherically symmetric space-times. This paper is outlined in
the following manner: In section \ref{sec2:action}, the action of
$f(R)$ gravity coupled to nonlinear electrodynamics is introduced,
and the respective gravitational field equations and
electromagnetic equations are presented. In section
\ref{sec3:simple}, we restrict the metric fields to one degree of
freedom, by considering the specific case of
$g_{tt}=-g_{rr}^{-1}$, and using the dual $P$ formalism of
nonlinear electrodynamics, we present exact solutions in terms of
the structural function $H_P$. Subsequently, in section
\ref{sec4:2dof} we investigate the situation where the two metric
fields are related via a power law in $r$, introducing additional
parameters, and derive new specific solutions. In section
\ref{sec5:alternative}, we present alternative methods of finding
exact solutions, first by considering the specific case of
constant curvature, then by choosing a form for the $f(R)$, before
we conclude in section \ref{sec6:conclusion}.

%-----------------------------------------------------------------
\section{Action and field equations}\label{sec2:action}
%-----------------------------------------------------------------

Throughout this work, we consider a static and spherically
symmetric space-time, in curvature coordinates, given by the
following line element
\begin{eqnarray}
 ds^2&=&-e^{2\alpha(r)}\,dt^2 +e^{2\beta(r)}\,dr^2+r^2\,
   (d\theta^2+\sin ^2{\theta}\, d\phi ^2)
 \,, \label{generalmetric}
\end{eqnarray}
where the metric fields $\alpha$ and $\beta$ are both arbitrary
functions of $r$. We use geometrized units, $c=G=1$.

The action describing $f(R)$ gravity coupled to nonlinear
electrodynamics is given in the following form
\begin{equation}
 S\ =\ \int \sqrt{-g}\left[\frac{f(R)}{2\kappa}+L(F)\right]\,d^4x \,,
\end{equation}
where $\kappa=8\pi$, and $f(R)$ is an arbitrary function of the
Ricci scalar $R$. $L(F)$ is a gauge-invariant electromagnetic
Lagrangian which depends on a single invariant $F$ given by
$F=F^{\mu\nu}F_{\mu\nu}/4$ \cite{Pleb}. As usual the antisymmetric
Faraday tensor $F_{\mu\nu}=A_{\nu,\mu}-A_{\mu,\nu}$ is the
electromagnetic field and $A_\mu$ its potential. In Maxwell theory
the Lagrangian takes the form $L(F)= -F/4\pi$. Nevertheless, we
consider more general choices of electromagnetic Lagrangians. The
Lagrangian may also be constructed using a second invariant $G
\sim F_{\mu\nu}{}^*F^{\mu\nu}$, where the asterisk $^*$ denotes
the Hodge dual with respect to $g_{\mu\nu}$. However, we shall
only consider $F$, as this provides interesting enough results.

%-----------------------------------------------------------------
\subsection{Gravitational field equations}\label{Sec2A:action}
%-----------------------------------------------------------------

Varying the action with respect to $g_{\mu\nu}$ provides the
following gravitational field equation
\begin{equation}
f_R R_{\mu\nu}-\frac{1}{2}f\,g_{\mu\nu}-\nabla_\mu \nabla_\nu
f_R+g_{\mu\nu}\Box f_R\ =\ \kappa \,T_{\mu\nu} \,,
    \label{2field:eq}
\end{equation}
where $f_R=df/dR$, and the stress-energy tensor of the nonlinear
electromagnetic field is given by
\begin{equation}
T_{\mu\nu}\ =\ g_{\mu\nu}\,L(F)-F_{\mu\alpha}F_{\nu}{}^{\alpha}
  \,L_{F}\,, \label{4dim-stress-energy}
\end{equation}
with $L_F=dL/dF$.

Taking into account the symmetries of the geometry given by the
metric (\ref{generalmetric}), the non-zero compatible terms for
the electromagnetic field tensor are
\begin{equation}
F_{\mu\nu}\ =\ 2E(x^\alpha)\,\delta^{[t}_\mu
\,\delta^{r]}_\nu+2B(x^\alpha)\,\delta^{[\theta}_\mu
\,\delta^{\phi]}_\nu
    \label{em-tensor}\,,
\end{equation}
such that the only non-zero components are $F_{tr}=E(x^\mu)$ and
$F_{\theta\phi}=B(x^\mu)$. Thus, the invariant $F$ takes the
following form
\begin{equation}
F\ =\ -\frac{1}{2}\left[e^{-2(\alpha+\beta)}\,E^2-\frac{B^2}{r^4
\sin^2\theta}\right]  \,.
    \label{invF}
\end{equation}
Consequently, the stress-energy tensor components are given by
\begin{eqnarray}
T^{t}{}_{t} &=\ T^{r}{}_{r} &=\ L+e^{-2(\alpha+\beta)}E^2\,L_F \,,
    \label{ttrr-rel}
    \\
T^{\theta}{}_{\theta} &=\ T^{\phi}{}_{\phi} & =\
L-\frac{B^2}{r^4\sin^2\theta}\,L_F \,.
    \label{stress:eq}
\end{eqnarray}
The property $T^{t}{}_{t}=T^{r}{}_{r}$ imposes a stringent
constraint on the field equations, which will be analyzed further
below.

The contraction of the field equation (\ref{2field:eq}) yields the
trace equation
\begin{equation}
f_R R-2f+3\,\Box f_R\ =\ \kappa T
    \label{2trace}
\end{equation}
which shows that the Ricci scalar is a fully dynamical degree of
freedom. The trace of the stress-energy tensor,
$T=T^{\mu}{}_{\mu}$, is given by $T=4(L-FL_F)$. Note that for the
Maxwell limit, with $L=-F/(4\pi)$ and $L(F)=-1/(4\pi)$, one
readily obtains $T=0$, and consequently Eq.~(\ref{2trace}) in the
Maxwell limit reduces to $f_R R-2f+3\,\Box f_R=0$.

The trace equation (\ref{2trace}) can be used to simplify the
field equations and then keep it as a constraint equation. Thus,
substituting the trace equation into the field equation
(\ref{2field:eq}), we end up with the following gravitational
field equation
\begin{equation}
f_R R^{\mu}{}_{\nu}-\frac{1}{4}\delta^{\mu}_{\nu}\left(f_R R -\Box
f_R -\kappa T\right)-\nabla^{\mu}\nabla_{\nu}f_R\ =\
\kappa\,T^{\mu}{}_{\nu}
    \,. \label{3field:eq}
\end{equation}
Now we can use the properties (\ref{ttrr-rel}) and
(\ref{stress:eq}) of the electromagnetic stress-energy tensor by
subtracting the $(rr)$--$(tt)$ and $(\theta\theta)$--$(tt)$
components, which provides the following field equations:
\begin{equation}
f_R^{\prime \prime }-\left(\alpha+\beta\right)'
f_R'-\frac{2}{r}\left(\alpha+\beta\right)'f_R\ =\ 0
\,,    \label{f1}
\end{equation}
and
\begin{eqnarray}
\frac{N(r)}{\kappa r^2}\ =\
-\left[e^{-2(\alpha+\beta)}E^2+\frac{B^2}{r^4\sin^2\theta}
\right]L_F  \,, \label{f2}
\end{eqnarray}
respectively, where we defined the dimensionless function $N(r)$ as
\begin{eqnarray}
N(r)\ =\ r^2e^{-2\beta}f_R\Bigg[\alpha''+2\alpha'^2
+\frac{e^{2\beta}-1}{r^2}+\left(\alpha'+\beta'
-\frac{f_R'}{f_R}\right)\left(\frac{1}{r}-\alpha'\right)\Bigg]
\,.  \label{def:N}
\end{eqnarray}
The prime stands for the derivative with respect to the radial
co-ordinate $r$. It is important to note that Eq.~(\ref{f1})
places a constraint on the metric fields and $f_R$, independently
of the form of the electromagnetic Lagrangian. In the Einstein
limit, $f_R=1$, Eq.~(\ref{f1}) leads to $(\alpha+\beta)'=0$ which
we will assume in section \ref{sec3:simple} to explore a specific
class of solutions.

Note that with help of Eq.~(\ref{f1}), the following
relationship
\begin{eqnarray}
\Box
f_R\ =\ e^{-2\beta}\left[f_R''+\left(\alpha'-\beta'
  +\frac{2}{r}\right)f_R'\right]\,,
\end{eqnarray}
and the definition of the curvature scalar, provided from the
metric, given by
\begin{equation}
R\ =\ 2e^{-2\beta}\left[\left(\alpha'+\frac{2}{r}\right)\left(\beta'
-\alpha'\right)-\alpha''+\frac{e^{2\beta}-1}{r^2}\right]\,,
\end{equation}
the trace equation (\ref{2trace}) may be expressed as
\begin{equation}
f\ =\ f_R e^{-2\beta}\left[-\alpha''+\alpha'(\beta'-\alpha')
+\frac{1}{r}(\alpha'+5\beta')+\frac{e^{2\beta}-1}{r^2}
+3\left(\alpha'+\frac{1}{r}\right)\frac{f_R'}{f_R}\right]
-\frac{\kappa}{2}T  \,. \label{f-trace}
\end{equation}
If $\alpha(r)$ and $\beta(r)$ are specified, one can obtain
$f_R(r)$ from the first gravitational equation (\ref{f1}) and the
curvature scalar in a parametric form, $R(r)$, from its definition
via the metric. Then, once $T$ is known as a function of $r$, one
may in principle obtain $f(R)$ as a function of $R$ from
Eq.~(\ref{f-trace}).

%-----------------------------------------------------------------
\subsection{Electromagnetic field equations: $F$ representation
of nonlinear electrodynamics}\label{Sec2B:action}
%-----------------------------------------------------------------

The electromagnetic field equations are given by the following
relationships
\begin{equation}
\left(F^{\mu\nu}\,L_{F}\right)_{;\mu}\ =\ 0 \;,  \qquad
\left(^*F^{\mu\nu}\right)_{;\mu}\ =\ 0 \,.
    \label{em-field}
\end{equation}
The first equation is obtained by varying the action with respect
to the electromagnetic potential $A_\mu$. The second relationship,
in turn, is deduced from the Bianchi identities.

Using the electromagnetic field equation
$\left(^*F^{\mu\nu}\right)_{;\mu}=0$, we obtain $E=E(r)$ and
$B=B(\theta)$, and from $\left(F^{\mu\nu}\,L_{F}\right)_{;\mu}=0$,
we deduce
\begin{equation}\label{ELF}
EL_F\ =\ \frac{q_e\,e^{(\alpha+\beta)}}{r^2} \,,  \qquad
B\ =\ q_m\sin\theta  \,.
\end{equation}

The electric field $E$ is determined from equations (\ref{f2}) and
(\ref{ELF}), and is given by
\begin{equation}\label{E-field}
E(r)\ =\ \frac{e^{\alpha+\beta}}{2\kappa q_e}\left[-N(r) \pm
\sqrt{N^2(r)-\left(\frac{2\kappa q_e q_m}{r^2}\right)^2}\right]
\,.
\end{equation}
Note that independently of $N(r)$ the electric field diverges at
the center in the presence of a magnetic field, as in the general
relativistic case \cite{Arellano3}. Thus, to avoid this problematic
feature, in the following analysis we consider either a purely
electric field or a purely magnetic field.

The physical fields and the other relevant quantities in the
purely electric and the purely magnetic case, respectively, are
summarized in the following table:
 \begin{equation} \label{pureE-field}
  \begin{array}[c]{r|c c c c}
   & & & & \\
   & E(r) & B(\theta) & F(r) &  L_F(r) \\
   & & & & \\
   \hline
   & & & & \\
   {\rm purely\ electric}\quad & \quad
 -e^{\alpha+\beta}\dfrac{N}{\kappa q_e} \quad  & 0 & \quad
 -\dfrac{1}{2}\left(\!\dfrac{N}{\kappa q_e}\!\right)^2 \quad &
 -\dfrac{\kappa q_e^2}{N}\dfrac{1}{r^2} \\
   & & & & \\
   \hline
   & & & & \\
   {\rm purely\ magnetic}\quad & 0 & \quad q_m\sin\theta \quad &
 \dfrac{q_m^2}{2}\dfrac{1}{r^4} & \quad -\dfrac{N}{\kappa
 q_m^2}r^2 \quad \\
   & & & &
  \end{array}
 \end{equation}
In the purely magnetic case the field equations assume a simpler
form, $Nr^2\propto L_F$, than in the purely electric case, where
$Nr^2\propto 1/L_F$, and the magnetic fields is independent of
the metric fields, contrary to the electric field. Therefore the
$F$ representation of electrodynamics is more suited for finding
purely magnetic solutions which, however, involve magnetic
monopoles.

%-----------------------------------------------------------------
\subsection{Electromagnetic field equations:
Dual $P$ formalism}\label{Sec2C:action}
%-----------------------------------------------------------------

As introduced above nonlinear electrodynamics is represented in
terms of a nonlinear electrodynamic field, $F_{\mu\nu}$, and its
invariants. However, one may introduce a dual representation in
terms of an auxiliary field $P_{\mu\nu}$. This strategy proved to
be extremely useful for deriving exact solutions in general
relativity, especially in the electric regime \cite{Garcia,Garcia2}.
The dual representation is obtained by the following Legendre
transformation
\begin{equation}\label{Hgen}
  H\ =\ 2FL_F-L\,.
\end{equation}
The structural function $H$ is a functional of the invariant $P=
P_{\mu\nu} P^{\mu\nu}/4$. Then the theory is recast in the $P$
representation by the following relations
\begin{eqnarray}\label{structP}
P_{\mu\nu}\ =\ L_F F_{\mu\nu}\,, \qquad
F_{\mu\nu}\ =\ H_P P_{\mu\nu} \,, \qquad
L\ =\ 2PH_P-H\,, \qquad
L_F H_P\ =\ (4\pi)^{-2}  \,,
\end{eqnarray}
where $H_P=dH/dP$. The invariant $P$ is given by
\begin{equation}\label{Pinv}
P\ =\ \frac{1}{4}\,P_{\mu\nu}P^{\mu\nu} \ =\
-\frac{1}{2}\left[e^{-2(\alpha+\beta)}P_{tr}^2
-\frac{1}{r^4\sin^2\theta}\,P_{\theta\phi}^2  \right] \,.
\end{equation}

The stress-energy tensor in the dual $P$ formalism is written as
\begin{equation}
T_{\mu\nu}\ =\
g_{\mu\nu}\,(2PH_P-H)-P_{\mu\alpha}P_{\nu}{}^{\alpha} \,H_{P}\,,
\end{equation}
and provides the following non-zero components
\begin{eqnarray}
T^{t}{}_{t} &=& T^{r}{}_{r}\ =\ -H+\frac{1}{r^4\sin^2\theta}\,
  P_{\theta\phi}^2\,H_P
  \,, \label{4TttP}   \\
T^{\theta}{}_{\theta} &=& T^{\phi}{}_{\phi}\ =\
-H-e^{-2(\alpha+\beta)} P_{tr}^2\,H_{P}
\,.  \label{4TppP}
\end{eqnarray}
The trace of the stress-energy tensor reads $T=-4(H-PH_P)$, so
that in the Maxwell limit, $H=-P/(4\pi)$ and $H_P=-1/(4\pi)$, we
have $T=0$, which is consistent with the $F$ formalism, as
outlined in Section \ref{Sec2A:action}.

The electromagnetic field equations now read
\begin{equation}
P^{\mu\nu}{}_{;\mu}\ =\ 0  \;,  \qquad \left(H_P
\,^*P^{\mu\nu}\right)_{;\mu}\ =\ 0 \,.
     \label{em-fieldP}
\end{equation}
We emphasize that the tensor $F_{\mu\nu}=H_P\,P_{\mu\nu}$ is the
physically relevant quantity. The $P$ invariant may be deduced
from Eqs.~(\ref{em-fieldP}) in an analogous manner as in the $F$
formalism. In the purely electric case, $B=0$ we find
\begin{equation}\label{P-E}
P\ =\ -\frac{q_e^2}{2r^4} \,.
\end{equation}
Due to the fact that it does not depend on the metric fields
$\alpha$ and $\beta$, this formalism is attractive to find
electric solutions, as opposed to the usual $F$ representation
where purely magnetic solutions are easier to find. The
gravitational field equation (\ref{f2}) now takes the simple form
\begin{equation}\label{fieldH}
r^2N(r)\ =\ -\kappa q_e^2 H_P(r) \,,
\end{equation}
where the function $N(r)$ was defined in Eq.~(\ref{def:N}) and
describes the gravity side. Through Eqs.~(\ref{pureE-field}) in
the purely electric case we can express the electric field in
terms of $H_P$ and $P$ as
\begin{equation}
 E(r)\ =\ \frac{q_e}{r^2} e^{\alpha+\beta}H_P
 \ =\ e^{\alpha+\beta}\sqrt{-2P}\,H_P \,. \label{geleralE-F}
\end{equation}

In summary, using the dual $P$ formalism, it is easier to find
nonlinear electrodynamic solutions than in the $F$ formalism, for
the specific case of pure electric fields. We shall consider
several specific solutions in the following section.

%-----------------------------------------------------------------
\section{Specific solutions: $\alpha(r)=-\beta(r)$}
\label{sec3:simple}
%-----------------------------------------------------------------

It is highly non-trivial to find general solutions for the field
equations of $f(R)$ modified theories of gravity coupled to
nonlinear electrodynamics. However, restricting the metric fields
to one degree of freedom provides very interesting solutions which
will be analyzed in this section. In this context, the condition
$(\alpha+\beta)'=0$ imposes $\alpha(r)=-\beta(r)$, where the
constant of integration can safely be absorbed by redefining the
time co-ordinate.

In this specific case Eq.~(\ref{f1}) implies $f_R(r)=Ar+B$. The
Einstein limit is achieved by $A\rightarrow 0, B\rightarrow 1$ and
so we define $\A:=A/B$ which represents the departure from
Einstein gravity while $B$ can be interpreted as rescaling the
coupling constants. The second field equation (\ref{fieldH}) now
provides the following general solution for the metric field in
terms of $H_P$:
\begin{eqnarray}
e^{2\alpha(r)}\ =\ 1 -\frac{2r^2}{3}\left( 3C_1 + \int
\Big[\frac{\kappa q_e^2}{B\bar{r}^2}H_P(\bar{r})-\A\bar{r}\Big]\,
\frac{\Gamma(\bar{r})}{\bar{r}^3}\,d\bar{r}\right)
+\frac{2\Gamma(r)}{3\,r}\left( \frac{C_2}{B} + \int
\Big[\frac{\kappa
q_e^2}{B\bar{r}^2}H_P(\bar{r})-\A\bar{r}\Big]\,d\bar{r} \right)
\,,
     \label{Maxwellmass}
\end{eqnarray}
where $C_1$ and $C_2$ are constants of integration and the
function $\Gamma(r)$ is defined as
\begin{equation}
\Gamma(r)\ =\ 1 -\frac{3}{2}\A r +3\A^2r^2
-3\A^3r^3\ln\!\Big(B[\A +1/r]\Big) \,.
\end{equation}
The electric field, given in Eq.~(\ref{geleralE-F}), in this case
simply provides
\begin{equation}
 E(r)\ =\ \frac{q_e}{r^2} H_P(r) \,.
\end{equation}

Thus, in principle, by choosing a particular nonlinear
electrodynamics theory, by specifying $H_P$, all the physical
fields are deduced. Note that in order for the electric field to
be finite at the center $H_P$ must be $\propto r^\varepsilon$ for
small $r$, with $\varepsilon \geq 2$. In the following sections
we consider specific choices for $H_P$ and find the respective
exact solutions.

%-----------------------------------------------------------------
\subsubsection{$f(R)$ gravity and Maxwell electrodynamics}
\label{sec:MaxGR}
%-----------------------------------------------------------------

Consider the specific case of $f(R)$ gravity coupled to Maxwell
electrodynamics, \ie{} $H(P)=-P/(4\pi)$. The field equation
(\ref{fieldH}) provides the following exact solution
\begin{equation}
e^{2\alpha(r)}\ =\ 1 +\A D -\frac{2D}{3r} +\frac{q_e^2}{Br^2}
-\big(1+2\A D\big)\A r -2C_1r^2 +\left[\frac{1}{2}+\big(1+2\A
D\big) \ln\!\Big(B[\A +1/r]\Big)\right]\A^2r^2
\,,\label{fRMaxwell}
\end{equation}
where we defined $D=(2\A\,q_e^2-C_2)/B$, which can be interpreted
as an effective mass for the $f(R)$-Maxwell case. The
corresponding electric field is simply
\begin{equation}
E(r)\ =\ -\frac{q_e}{4\pi r^2} \,,
\end{equation}
and, as expected, diverges at the center.

Note that the vacuum solution, $H(P)=0$, in $f(R)$ gravity, can
be immediately obtained by setting $q_e=0$ in the Maxwell
solution, Eq.~(\ref{fRMaxwell}),
\begin{equation}
e^{2\alpha(r)}\ =\ 1 -\frac{C_2\A}{B} +\frac{2C_2}{3Br}
-\Big(1-\frac{2C_2\A}{B}\Big)\A r -2C_1r^2
+\left[\frac{1}{2}+\Big(1-\frac{2C_2\A}{B}\Big)
\ln\!\Big(B[\A +1/r]\Big)\right]\A^2r^2 \,.
\end{equation}
An interesting difference to the vacuum solution in general
relativity is the term linear in $r$, and the term with the
logarithm. Note that the former linear term also arises in the
vacuum solutions of conformal Weyl gravity \cite{MannKaz}.

In order to obtain the Schwarzschild-de Sitter solution, one sets
the following values for the constants: $\A=0$, $C_2=-3BM$ and
$C_1=\Lambda/6$. This result is similar to the analysis outlined
in Ref.~\cite{Multamaki:2006zb}. Note also that the $\A\neq 0$
vacuum solution is not asymptotically flat. An interesting
solution is obtained by setting $C_2=0$, which yields
\begin{equation}
e^{2\alpha(r)}\ =\ 1 -\A r -2C_1r^2 +\left[\frac{1}{2}
+\ln\!\Big(B[\A +1/r]\Big)\right]\A^2r^2 \,.
\end{equation}
This solution has no effective mass term. For positive $B$ it is
regular at the center but diverges for large $r$, independently of
the constants $C_1$ and $\A$. For negative $B$ it shows the
opposite behavior.

For the specific case of general relativity coupled to Maxwell
electrodynamics, \ie{} $f(R)=R$ and $H(P)=-P/(4\pi)$, the solution
reduces to
\begin{equation}
e^{2\alpha(r)}\ =\ 1+\frac{2C_2}{3r}+\frac{q_e^2}{r^2}-2C_1r^2
   \label{sol5}
\end{equation}
which is simply the Reissner-Nordstrom-de Sitter solution by
setting $C_2=-3M$ and $C_1=\Lambda/6$, as shown above.
Note that the solution (\ref{sol5}) is equivalent to considering
$A=0$ and $B=1$ in the solution given by
Eq.~(\ref{fRMaxwell}).

Clearly it is interesting to try to reconstruct the $f(R)$ theory
associated with the solution given in Eq.~(\ref{fRMaxwell}). First
we calculate the Ricci scalar for the given $\alpha(r)$ which reads
in parametric form
\begin{eqnarray}
& R(r)\ =\ \dfrac{1}{(1+\A r)^2}\Bigg[ 24C_1 +13\A^2 +36\A^3 D
-\dfrac{2\A D}{r^2} +\dfrac{6\A +8\A^2 D}{r} +24\left(2C_1
+\A^3 D\right)\A r &
  \nonumber \\
& +6\left(4C_1-\A^2\right)\A^2r^2 -12\A^2 (1+2\A D)
(1+\A r)^2\ln\Big(B[\A + 1/r]\Big) \Bigg] \,. &
\end{eqnarray}
Because of the term $\propto\ln(\A+1/r)$, however, this cannot
simply be inverted to find $r(R)$. Using the trace equation
(\ref{f-trace}), with $T=0$ for the Maxwell case, we find $f(r)$
in parametric form
\begin{eqnarray}
& f(r)\ =\ \dfrac{B}{2(1+\A r)^2}\Bigg[ \A(9+24C_1+13\A^2
+36\A^3 D) +\dfrac{6+6\A^2+8\A^3 D}{r}
-2D\dfrac{1+\A^2}{r^2} +6\left(4\A^3 C_1-\A^5\right)r^2 &
  \nonumber \\
& +4\A^2\left(1+12C_1+6\A^3 D\right)r  -12\A^3\left(1 +2\A
D\right)\left(1+\A r\right)^2 \ln\Big(B \left[ \A +1/r\right]
\Big) \Bigg] \,. &   \label{f_from_trace}
\end{eqnarray}
In principle one could find the functional form $f(R)$ from these
parametric forms but, as mentioned, $R(r)$ can not analytically be
inverted to find $r(R)$ and substitute into $f(r)$.

\begin{figure}[!ht]
  \begin{center}
    \includegraphics[width=0.45\textwidth]{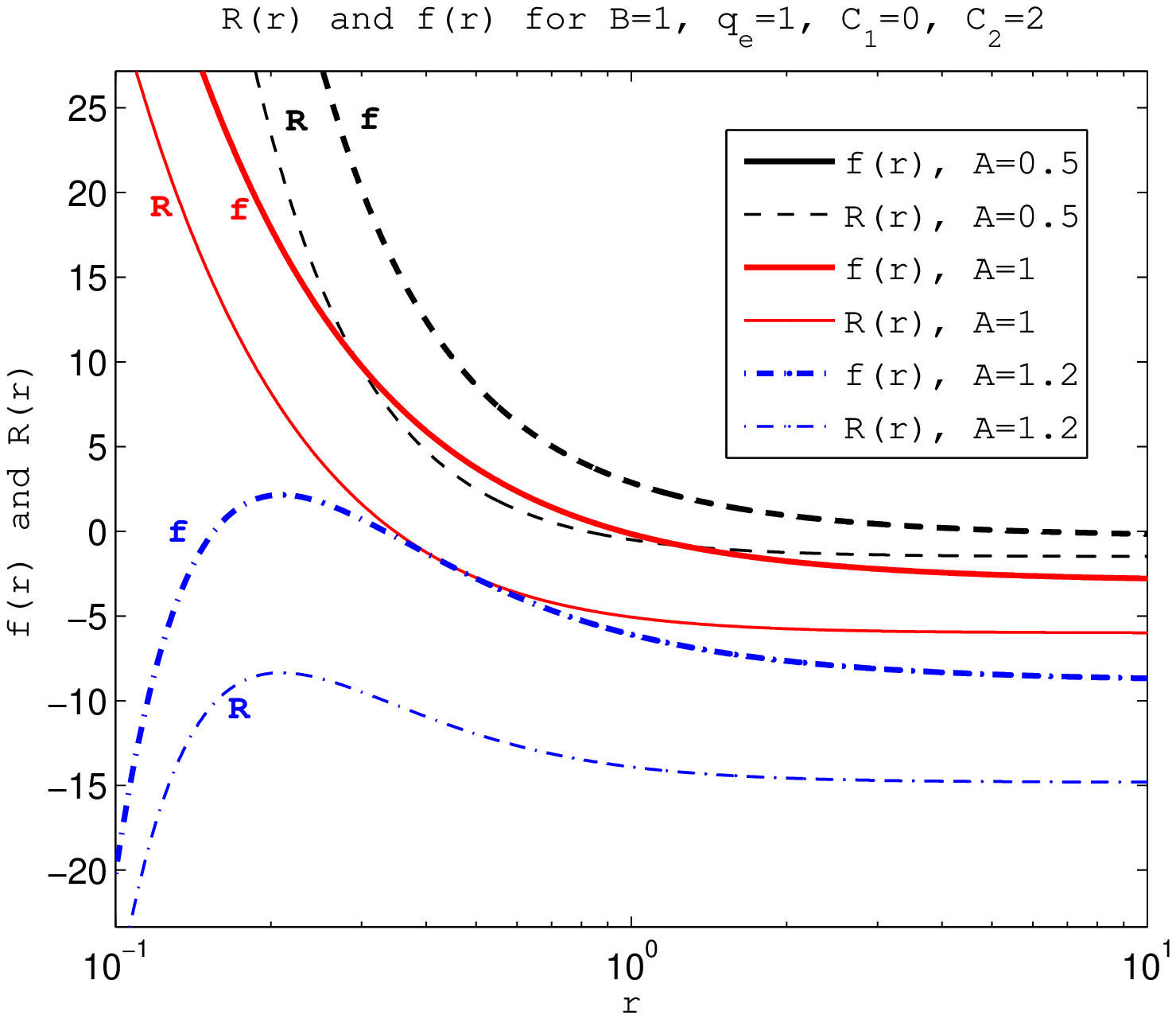}
    \qquad
    \includegraphics[width=0.45\textwidth]{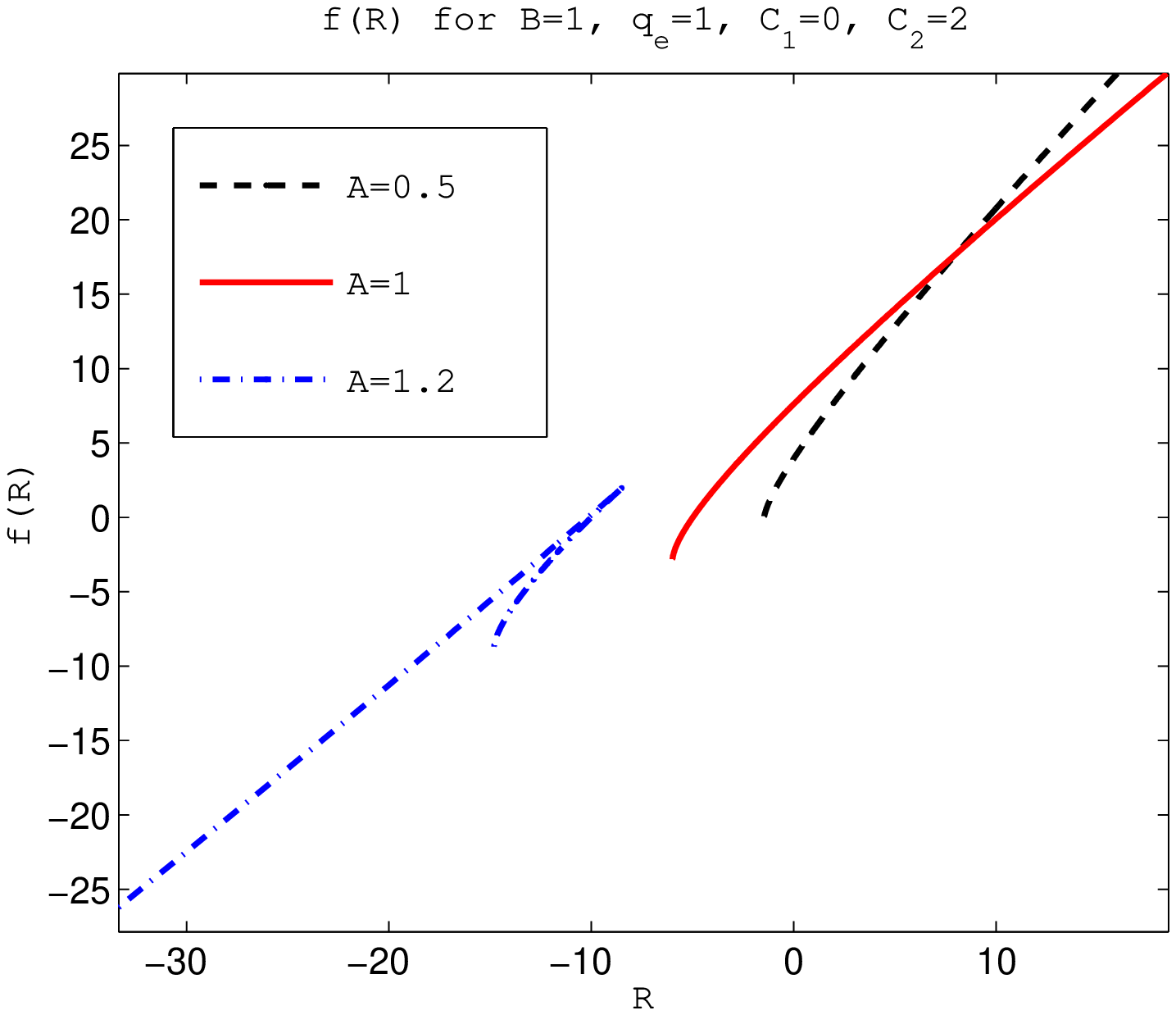}
      \caption{In the left panel we plot $R(r)$ and $f(r)$, while
      the right panel shows $f(R)$. We choose the coupling
      constants to be $B=1=q_e$ and the overall additive constant
      $C_1=0$. The sign of the constant $D$ influences the sign
      of $R$ and $f$ close to the center. For the choice $C_2=2$
      it reduces to $D=2(A-1)$ and by plotting $f$ and $R$ for
      $A\in\{0.5, 1, 1.2\}$  we see the different behaviors: as
      soon as $D$ is positive (here $A>1$) $f(R)$ is not a
      uniquely defined function at large distances $r$ from the
      center.}
    \label{fig:fandR}
  \end{center}
\end{figure}

In figure \ref{fig:fandR} we plot $R(r)$, $f(r)$, and $f(R)$ for
specific values of the constants. $B$ and $q_e$ scale the
gravitational and electromagnetic force while $C_1$ only acts as
an overall additive constant, so we set $B=1=q_e$ and $C_1=0$ in
the plots. The sign of the constant $D=(2\A q_e^2-C_2)/B$
influences the sign of $R$ and $f$ close to the center. Thus by
looking at different values of $A$ with fixed $C_2=2$, we see the
different behaviors: if $D$ is positive $f(R)$ is not a uniquely
defined function at large distances $r$ from the center. As a
consequence, for $f(R)$ to be well-defined everywhere $C_2$ needs
to satisfy $C_2\geq 2\A q_e^2$.

%-----------------------------------------------------------------
\subsubsection{Generalized Maxwell electrodynamics}
\label{sec:Maxlimit}
%-----------------------------------------------------------------

To demonstrate the effect of nonlinear electrodynamics we consider
a generalized form of the Maxwell theory described by the
following structural function
\begin{equation}
H(P)\ =\ -\frac{P}{4\pi}\left[1+\frac{\mu}{1+\delta}
(-2q_e^2P)^\delta\right] \,,
   \label{Lmaxlimit}
\end{equation}
where $\mu$ and $\delta$ are the characteristic parameters of the
theory. Note that this choice may physically describe strong
fields, as the second term is now dominant, \ie{} for $P\gg 1$.
This Lagrangian possesses the correct Maxwell limit for
$\delta>0$, \ie{} $H\simeq -P/(4\pi)$ for $P\ll 1$. The relevant
quantity $H_P$ is then given by
\begin{equation}
H_P\ =\ -\frac{1}{4\pi}\left[1+\mu(-2q_e^2P)^\delta\right] \,.
   \label{LFmaxlimit}
\end{equation}

A particularly interesting and simple example is obtained by
setting $\delta=1/4$, such that using Eq.~(\ref{P-E}), $H_P$ takes
the form
\begin{equation}
H_P\ =\ -\frac{1}{4\pi}\left(1+\frac{\mu q_e}{r}\right)\,.
   \label{LFmaxlimit2}
\end{equation}

Thus, substituting Eq.~(\ref{LFmaxlimit2}) into
Eq.~(\ref{fieldH}), we finally deduce the following solution:
\begin{eqnarray}
e^{2\alpha(r)} &=& 1 +\A\D -\frac{2\D}{3r}
+\frac{q_e^2(2-\A|q_e|\mu)}{2Br^2} +\frac{2|q_e|^3\mu}{5Br^3}
\nonumber \\
&& -\big(1+2\A\D\big)\A r -2C_1r^2
+\left[\frac{1}{2}+\big(1+2\A\D\big) \ln\!\Big(B[\A
+1/r]\Big)\right]\A^2r^2 \,,
   \label{Maxwellmass:sol}
\end{eqnarray}
where now the effective mass is generalized to
\begin{equation}
\D\ =\ \frac{\A\,q_e^2(2-\A|q_e|\mu)-C_2}{B} \,.
\end{equation}
Note that $f(R)$ gravity coupled with Maxwell electromagnetism,
\ie{} $H_P=-1/(4\pi)$, follows from the above solution in the
limit of $\mu=0$, which simply reduces to Eq.~(\ref{fRMaxwell}).
Note that we can use the solution (\ref{Maxwellmass:sol}) to
write $R(r)$ and $f(r)$ in parametric form, and finally, in
principle, deduce the functional form $f(R)$. However, as outlined
in Section \ref{sec:MaxGR}, $R(r)$ cannot be analytically inverted
to find $r(R)$ and substitute into $f(r)$. In addition to this, we do not write
out the explicit forms of $R(r)$ and $f(r)$ due to their lengthy
character.

Setting $A=0$ and $B=1$, which is equivalent to general
relativity, Eq.~(\ref{Maxwellmass:sol}) provides a particularly
interesting solution given by
\begin{equation}
e^{2\alpha(r)}\ =\ 1+\frac{2C_2}{3r}+\frac{q_e^2}{r^2}
+\frac{2|q_e|^3\mu}{5r^3}-2C_1r^2 \,,
\end{equation}
which can also be found from Eq.~(\ref{Maxwellmass}). Note the
presence of a term proportional to $1/r^3$, which dominates for
low values of $r$. This solution tends to the Maxwell-Einstein
limit setting $\mu=0$.

%-----------------------------------------------------------------
\section{New solutions: $\alpha(r) \neq -\beta(r)$} \label{sec4:2dof}
%-----------------------------------------------------------------

Due to the fact that $f(R)$ gravity has more degrees of
freedom compared to Einstein gravity, and also in view of Ref.
\cite{Jacobson:2007tj}, it is very interesting to explore the
situation of $\alpha\neq-\beta$. However, without specifying a
relation between $\alpha$ and $\beta$, a specific nonlinear
electrodynamics model, or a specific $f(R)$ theory the equations
are not closed and therefore analytically intractable. In this
section we consider the specific example where the two metric
fields satisfy the following relationship
\begin{equation}
 \alpha(r) + \beta(r)\ =\ \ln( kr^\ell )
\end{equation}
where $k$ and $\ell$ are free parameters. This case is
particularly interesting since it allows for regular electric
fields at the center, as will be shown below.

\begin{figure}[!ht]
  \begin{center}
    \includegraphics[width=0.45\textwidth]{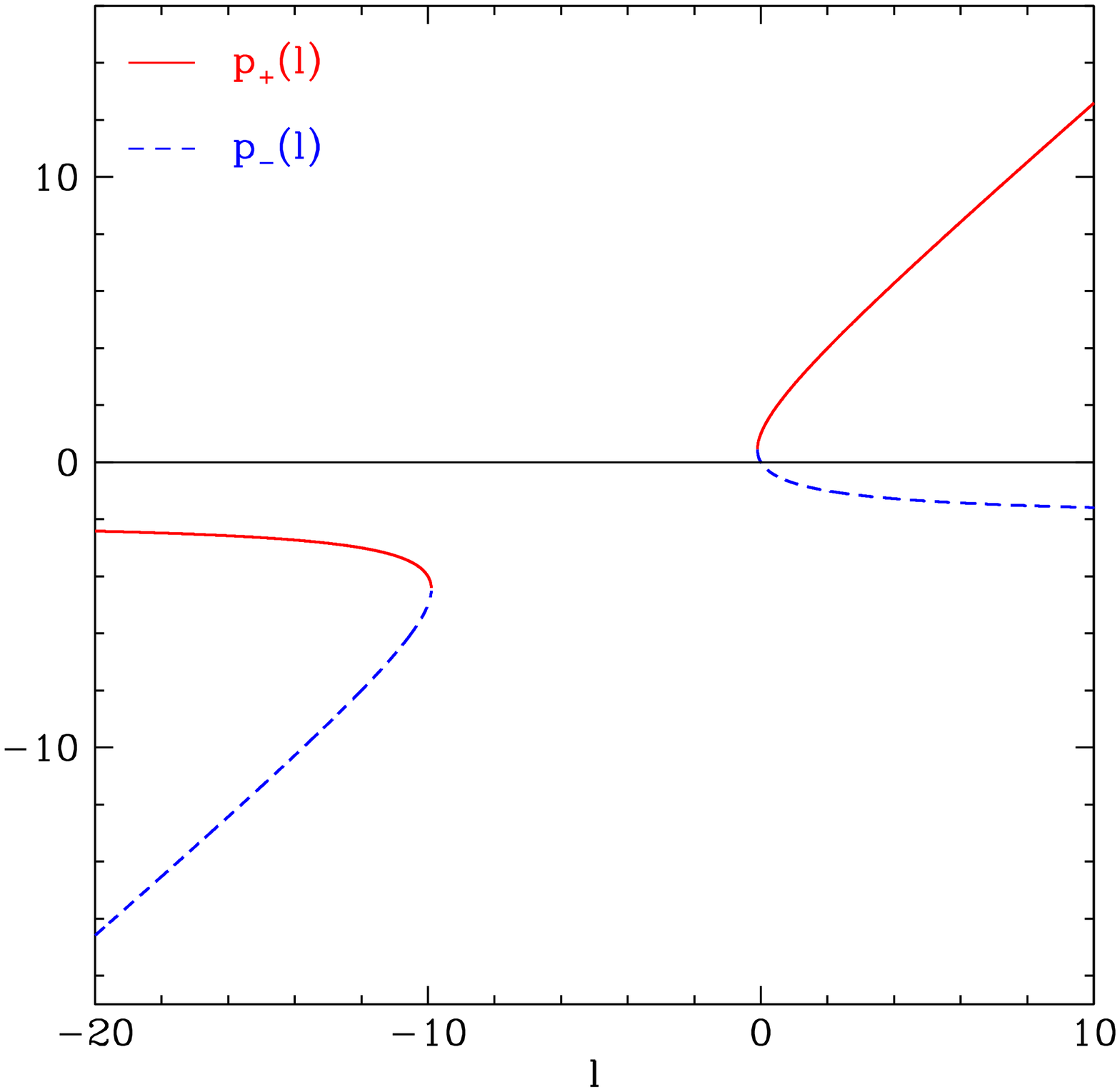}
    \qquad
    \includegraphics[width=0.45\textwidth]{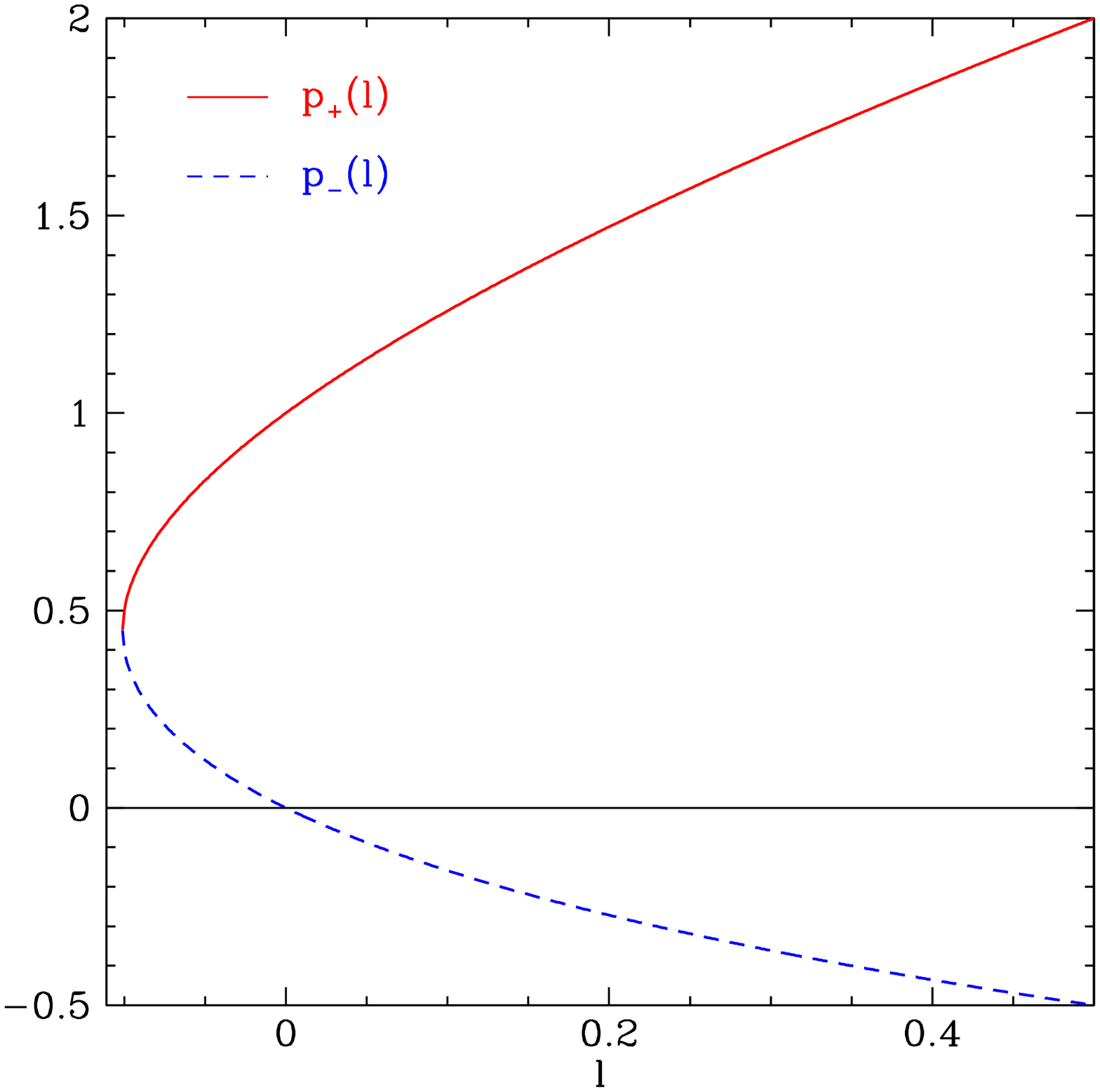}
      \caption{In the left panel we plot $\ppm$, the powers of
      $r$ in the solution for $f_R(r)$, as function
      of the parameter $\ell$. The right panel zooms into the
      right branch of $\ppm$. Note that $\pmin$ vanishes for
      $\ell=0$, is positive for negative $\ell$ and vice versa,
      while $\pmin\rightarrow-2$ for large $\ell$. This behavior
      makes the solution $f_R=Br^{\pmin}$ particularly interesting.}
    \label{fig:Ppm}
  \end{center}
\end{figure}

From the first field equation (\ref{f1}) we find that $f_R(r)$ has
the following form
\begin{equation}
 f_R(r)\ =\ Ar^{\pplus(\ell)} + Br^{\pmin(\ell)} \,,
\end{equation}
where $A$ and $B$ are constants of integration, and the exponents
depend on the parameter $\ell$ as
\begin{equation}
 \ppm(\ell)\ =\ \frac{1}{2}\left( 1
    + \ell \pm \sqrt{\ell\,(\ell+10)+1} \right) \,.
    \label{def:p}
\end{equation}
In order for $\ppm(\ell)$ to be real it is required that either
$\ell\geq 2\sqrt{6}-5$ or $\ell\leq -2\sqrt{6}-5$. In the limit
$\ell=0$ the exponents become $\pplus(0)=1$ and $\pmin(0)=0$
such that $f_R(r)=Ar+B$ as in the case considered in the previous
section. We plot $\ppm$ in figure \ref{fig:Ppm}.

The electric field now reads
\begin{equation}
 E(r)\ =\ q_e k r^{\ell-2} \,H_P(r) \,.
\end{equation}
An interesting case is $\ell=2$ where the electric field is
constant in the Maxwell limit, $H_P=-1/(4\pi)$.

Let us consider the case $A=0$ where $f_R(r)$ is a simple power
law which has a well defined Einstein limit for $\ell=0$ and
$B=1$. For this specific case the second field equation
(\ref{fieldH}) can in principle be solved for any general
structural function $H_P(r)$. We define the constant $K=\kappa
q_e^2 k^2/B$ and solve equation (\ref{fieldH}) for $\ell\neq 1,
2$, which provides the following solution
\begin{equation}
 e^{2\alpha}\ =\ \frac{k^2\,r^{2\ell}}{(\ell-1)(\pplus-2\ell-2)}
 + \frac{2\,r^2}{\pplus-4} \left[ \Big(C_1 + K \int\bar{r}^{\pplus
 +\ell-6}H_P(\bar{r})d\bar{r} \Big) - r^{\pplus-4}\Big(C_2
 +K\int\bar{r}^{\ell-2}H_P(\bar{r})d\bar{r}\Big) \right] \,,
\end{equation}
while the special cases $\ell=1,2$ have to be solved separately.
For $\ell=1$, the solution is
\begin{equation}
 e^{2\alpha}\ =\ \frac{\left(6\ln(r)-\sqrt{3}-3\right)k^{2}\,r^2}{3(\sqrt{3}-3)}
 + \frac{2\,r^2}{3(\sqrt{3}-3)} \left[ \Big(C_1 + K \int\bar{r}^{\sqrt{3}-4}
 H_P(\bar{r})d\bar{r} \Big) - r^{\sqrt{3}-3}\Big( C_2 +
 K\int\frac{H_P(\bar{r})}{\bar{r}} d\bar{r}\Big) \right] \,,
\end{equation}
and for $\ell=2$, we find
\begin{equation}
 e^{2\alpha}\ =\ -\frac{1}{2}k^{2}\,r^4 + 2\,r^2 \left[ \Big(C_1 +
 K \int \ln(\bar{r}) H_P(\bar{r}) d\bar{r} \Big) - \ln(r)\Big( C_2
 + K \int H_P(\bar{r}) d\bar{r} \Big) \right] \,.
\end{equation}
In all cases, the solution is not conformally flat due to the
first term.

An interesting case is $f(R)$ gravity coupled to Maxwell
electrodynamics where $H_P=-1/(4\pi)$. The electric field for
this case is given by
\begin{equation}
 E(r)\ =\ -\frac{q_e k}{4\pi} r^{\ell-2}
 \label{eq:Enew}
\end{equation}
where for $\ell=0$, the classical Coulomb field is recovered. For
$\ell<2$ it diverges at the center. Interestingly the electric
field is constant for $\ell=2$, as mentioned before. For $\ell>2$
the electric field vanishes at the center and diverges at spatial
infinity.

\begin{figure}[!ht]
  \begin{center}
    \includegraphics[width=0.45\textwidth]{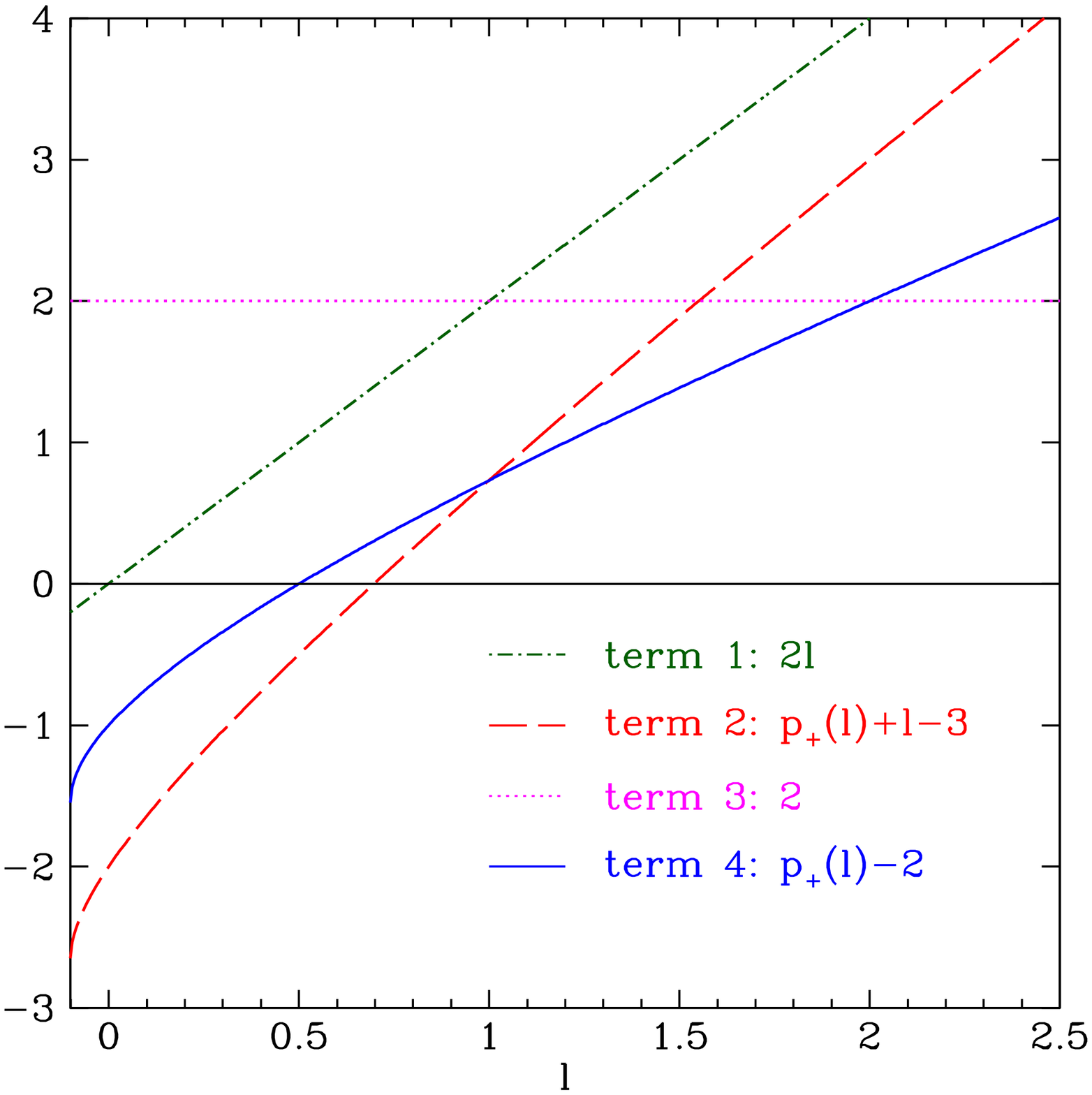}
    \qquad
    \includegraphics[width=0.45\textwidth]{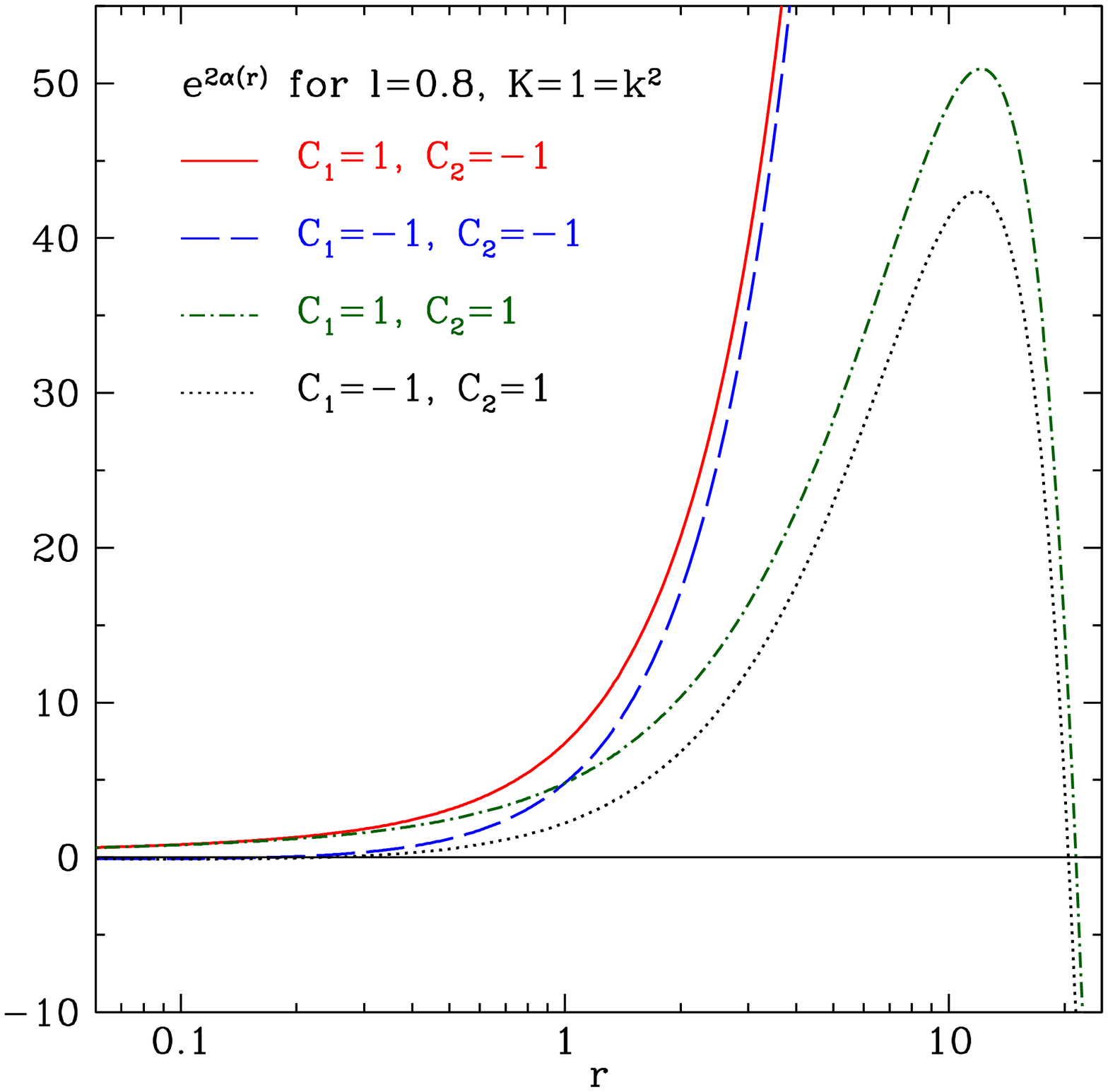}
      \caption{In the left panel the exponents of $r$ in the
      four terms of $e^{2\alpha(r)}$ for $\ell\neq 1, 2$ are
      plotted. For $\ell\geq \ell_{\rm crit}\simeq 0.7$ the
      exponents of $r$ are positive in all terms.
      Negative $\ell$ lead to
      solutions divergent at the center. At $\ell=1$ and $\ell=2$
      the hierarchy of the terms change which explains why these
      are special cases. The right panel shows $e^{2\alpha(r)}$
      for $\ell=0.8$ for four different combinations of signs of
      the integration constants, while for simplicity and
      transparency we set $K=1=k^2$ throughout the analysis.}
    \label{fig:powers}
  \end{center}
\end{figure}

The metric field is then given in the three cases as:
\begin{eqnarray}
 \ell \neq 1, 2\ : & \qquad & e^{2\alpha}\ =\ \frac{k^2\,r^{2\ell}}
 {(\ell-1)(\pplus-2\ell-2)}
   +\frac{K\,r^{\pplus+\ell-3}}{2\pi(\ell-1)(\pplus+\ell-5)}
   +\frac{2\left( C_1\,r^2 + C_2\,r^{\pplus-2} \right)}{\pplus-4} \,,
   \label{Eql1} \\
 \ell = 1\ : & \qquad & e^{2\alpha}\ =\ \frac{3-\sqrt{3}}{3(2-\sqrt{3})}
 \left\{ -C_1\,r^2 +C_2\,r^{\sqrt{3}-1} -\frac{K}{4\pi}\left[ \ln(r)
 +(3-\sqrt{3})^{-1} \right]r^{\sqrt{3}-1} \right\} \,,
\label{Eql2} \\
 \ell = 2\ : & \qquad & e^{2\alpha}\ =\ -\frac{1}{2}k^2\,r^4 +
 \frac{K}{2\pi}\,r^3 + 2\,r^2 \left[C_1-C_2\ln(r)\right] \,.
 \label{Eql3}
\end{eqnarray}
In the first case, the exponents of $r$ are positive in all terms
for $\ell\geq \ell_{\rm crit}$, where $\ell_{\rm
crit}=(5-\sqrt{13})/2\simeq 0.7$. See the left panel of figure
\ref{fig:powers} for a comparison of the different exponents. At
$\ell=1$ and $\ell=2$ the hierarchy of the terms change which
explains why these are special cases.

A solution is regular at the origin if the function and all its
derivatives are finite at $r=0$. We verify that the solutions
(\ref{Eql2}) and (\ref{Eql3}) are not regular at the origin,
although they vanish for $r=0$. In case of solution (\ref{Eql1})
we only consider $2<\ell\in\mathbb{N}$ for which the electric
field is regular at the origin, \cf{}Eq.~(\ref{eq:Enew}). For its
derivatives to be finite at the origin the following exponents of
$r$
\begin{eqnarray}
n_1 \ =& p_+ + \ell - 3
    & =\ \frac{1}{2}\left[ 3\ell -5
    +\sqrt{\ell \left(\ell+10\right)+1} \right] \,, \label{n2}
\\
n_2 \ =& p_+ - 2
    & =\ \frac{1}{2}\left[ \ell -3
    +\sqrt{\ell \left(\ell+10\right)+1} \right] \label{n3}
\end{eqnarray}
must be natural numbers, \ie{} $n_1, n_2 \in\mathbb{N}$.

The metric function $g_{rr}$ must also be regular at the origin.
Using $\alpha(r) + \beta(r)\ =\ \ln( kr^\ell )$, we have
\begin{equation}
e^{2\beta}\ =\ \frac{k^2}{\overline{C}_1+\overline{C}_2r^{p_+-l-3}
+\overline{C}_3\left(C_1r^{2(1-l)}+C_2r^{p_+-2-2l}\right)} \,,
\end{equation}
with the constants
\begin{equation}
\overline{C}_1\ =\ \frac{k^2}{(\ell-1)(\pplus-2\ell-2)}\,,\qquad
\overline{C}_2\ =\ \frac{K}{2\pi(\ell-1)(\pplus+\ell-5)}\,,\qquad
\overline{C}_3\ =\ \frac{2}{\pplus-4}
\end{equation}
We have considered that $l>2$, so that we have to impose
$C_1=0$ to have regularity at the origin. Furthermore the metric
function $e^{2\beta}$ and its derivatives need to exist at the
origin which means the following exponents of $r$ must be natural
numbers
\begin{eqnarray}
m_1 \ =& p_+ - \ell - 3
    & =\ \frac{1}{2}\left[ -\ell -5
    +\sqrt{\ell \left(\ell+10\right)+1} \right] \,, \label{def:m1}
\\
m_2 \ =& p_+ - 2(\ell+1)
    & =\ \frac{1}{2}\left[ -3(\ell +1)
    +\sqrt{\ell \left(\ell+10\right)+1} \right] \,. \label{def:m2}
\end{eqnarray}
However, it turns out that $m_1$ and $m_2$ are both negative for
all $\ell\in\mathbb{N}$ which contradicts the imposition that
$m_1$ and $m_2$ be natural numbers. Thus, we conclude that the
solution (\ref{Eql1}) is not regular at the origin.

It is evident that for $\ell=1$ and $\ell=2$ the metric field
goes to negative infinity for large $r>r_0$ and thus has to be
matched to an external vacuum solution at a junction interface at
$r<r_0$. This behavior is independent of the signs of the
constants of integration $C_1$ and $C_2$. However, for the
solution to be positive for small $r$, we find $C_2\geq 0$. In
the case of $\ell>2$ we find the same behavior of the metric
field and the same constraint on $C_2$. For $\ell_{\rm
crit}\leq\ell<1$ we find positive solutions for all $r$ if
$C_2\leq 0$ and $C_1\geq0$. If $C_1<0$ the solution is negative
for small $r>0$, if $C_2>0$ it tends to negative infinity for
large $r$. See right panel of figure \ref{fig:powers}.

We also emphasize that through $R(r)$ and $f(r)$ (not written
out explicitly due to their extremely lengthy nature) expressed in
parametric form, the functional form $f(R)$ may in principle be
deduced. However, as outlined in Section \ref{sec:MaxGR}, $R(r)$
cannot be analytically inverted to find $r(R)$ and substitute into
$f(r)$.

%-----------------------------------------------------------------
\section{Alternative approaches}\label{sec5:alternative}
%-----------------------------------------------------------------

%-----------------------------------------------------------------
\subsection{Constant curvature}\label{sec:constR}
%-----------------------------------------------------------------

An interesting alternative is to consider the specific case of
constant curvature $R(r)=R_0$. Note that in this case $f_R$ is
independent of $r$, and for simplicity one may set $f_R(r)=B$.
Thus, one verifies that Eq.~(\ref{f1}) yields
$\alpha(r)=-\beta(r)$, so that the curvature scalar is given by
\begin{equation}\label{Ricciscalar}
R\ =\ -\frac{2e^{2\alpha(r)}}{r^2}\left[4r\alpha'+r^2\alpha''
+2r^2(\alpha')^2-e^{-2\alpha}+1\right] \,.
\end{equation}
For constant curvature, $R(r)=R_0$, this yields the following
solution for $\alpha(r)$
\begin{equation}
e^{2\alpha(r)}\ =\ 1-\frac{2C_1}{r}+\frac{2C_2}{r^2}
-\frac{R_0}{12}r^2\,.
   \label{metric4}
\end{equation}

Substituting the metric field into Eq.~(\ref{fieldH}), one deduces
$H_P$, given by
\begin{equation}
H_P(r)\ =\ -\frac{C_2B}{2\pi q_e^2}\,.
\end{equation}
which reduces to the Maxwell type, \ie{} $H_P=-1/(4\pi)$, by
setting the constant of integration $C_2=q_e^2/(2B)$.

For this case, \ie{} constant curvature, and taking into account
that the Maxwell limit implies $T=0$ (see Section
\ref{Sec2A:action}), the trace equation (\ref{2trace}) imposes the
following algebraic relationship
\begin{equation}
f_R R-2f\ =\ 0 \,,
\end{equation}
so that the form of $f(R)$ needs to obey this algebraic identity.
Thus the metric given by Eq.~(\ref{metric4}) is an exact solution
for the class of solutions $f(R)$, in the Maxwell limit, that
satisfy $f_R(R_0) R_0-2f(R_0)=0$. For instance, considering the
case of $f(R)=R-\mu^4/R$, and using the above trace equation
yields $R_0=\pm \sqrt{\mu^2}$. The case of $f(R)=R+\gamma^2 R^2$,
provides $R_0=0$.

%-----------------------------------------------------------------
\subsection{Specific gravity theory: $f(R)=R+\bar{\gamma}^2 R^2$}
\label{sec:specificfR}
%-----------------------------------------------------------------

Another alternative approach is to consider specific choices for
the form of $f(R)$. Consider the specific case of
$f(R)=R+\bar{\gamma}^2 R^2$, which for $\alpha(r)=-\beta(r)$
implies that $Ar+B=1+\gamma^2 R$, with $\gamma^2=2\bar{\gamma}^2$.
Substituting the value for $R(r)$, provides the following
solution:
\begin{equation}
e^{2\alpha(r)}\ =\ 1-\frac{2C_2}{r}+\frac{2C_1}{r^2}
+\frac{(1-B)}{12\gamma^2}r^2 -\frac{A}{20\gamma^2}r^3 \,.
\end{equation}
Note that this solution is consistent with Eq.~(\ref{f1}), \ie{}
$f_R''(r)=\gamma^2R''(r)=0$.

Now, substituting this solution in Eq.~(\ref{f2}), and finally
using the relationship $P=-q_e^2/(2r^4)$, we reconstruct the
following nonlinear electrodynamic structural function
\begin{eqnarray}
H(P)&=&-\frac{C_1B}{2\pi q_e^2}P+\frac{A}{2\pi q_e^2}
\Bigg[3C_2|q_e|(-2P)^\frac{1}{2}-2|q_e|^\frac{3}{2}(-2P)^\frac{1}{4}
+\frac{B|q_e|^\frac{5}{2}}{5\gamma^2}(-2P)^{-\frac{1}{4}}
+\frac{A|q_e|^\frac{3}{2}}{8\gamma^2}(-2P)^{-\frac{1}{2}}\Bigg]\,.
\end{eqnarray}
Note that for $A=0$ and $C_1=q_e^2/(2B)$ it reduces to the Maxwell
type, \ie{} $H=-P/(4\pi)$. However, for $A\neq 0$ this structural
function does not tend to the Maxwell limit for $P\ll 1$.
Therefore it is not a viable nonlinear electrodynamic theory. This
specific case illustrates the difficulty in finding viable
nonlinear electrodynamic theories, \ie{} with the correct Maxwell
limit, by explicitly providing a form for $f(R)$.

%-----------------------------------------------------------------
\section{Conclusion}\label{sec6:conclusion}
%-----------------------------------------------------------------

The issue of exact static and spherically symmetric solutions in
$f(R)$ modified theories of gravity is an important theme, mainly
due to the analysis of weak field solar system constraints, and
the generalization of exact general relativistic solutions to
$f(R)$ gravity. In this work we have analyzed exact solutions of
static and spherically symmetric space-times in $f(R)$ modified
theories of gravity coupled to nonlinear electrodynamics. Firstly,
the metric fields were restricted to one degree of freedom, by
considering the specific case of $g_{tt}=-g_{rr}^{-1}$. Using the
dual $P$ formalism of nonlinear electrodynamics, an exact general
solution was found in terms of the structural function $H_P$. In
particular, exact solutions to the gravitational field equations
were found, confirming previous results and new pure electric
field solutions were deduced. Secondly, by allowing two degrees of
freedom for the metric fields, and motivated by the existence of
regular electric fields at the center, new solutions were found.
Finally, we have also briefly considered alternative approaches by
analyzing the specific case of constant curvature and secondly, by
considering a specific form for $f(R)$.

%-----------------------------------------------------------------
\acknowledgments
The authors  thank an anonymous referee for very constructive
comments and suggestions. LH thanks Robert Crittenden for advice
and support. FSNL acknowledges funding by Funda\c{c}\~{a}o para
a Ci\^{e}ncia e a Tecnologia (FCT)--Portugal through the grant
SFRH/BPD/26269/2006.

%-----------------------------------------------------------------


\begin{thebibliography}{99}

\bibitem{expansion}
S.~Perlmutter {\it et al.}, Astrophys.\ J.\  {\bf 517}, 565 (1999);
A.~G.~Riess {\it et al.}, Astron.\ J.\  {\bf 116}, 1009 (1998);
A.~G.~Riess {\it et al.}, Astrophys.\ J.\  {\bf 607}, 665 (2004)
A. Grant {\it et al}, Astrophys. J. {\bf 560} 49-71 (2001);
S. Perlmutter, M. S. Turner and M. White,
 Phys. Rev. Lett. {\bf 83} 670-673 (1999);
C. L. Bennett {\it et al}, Astrophys. J. Suppl. {\bf 148} 1 (2003);
G. Hinshaw {\it et al}, Astrophys.\ J.\ Suppl.\  {\bf 148}, 135 (2003).

\bibitem{Starobinsky:1980te}
A.~A.~Starobinsky, Phys.\ Lett.\  B {\bf 91}, 99 (1980).

\bibitem{Carroll:2003wy}
S.~M.~Carroll, V.~Duvvuri, M.~Trodden and M.~S.~Turner, Phys.\
Rev.\ D {\bf 70}, 043528 (2004).
%[arXiv:astro-ph/0306438].

\bibitem{Fay:2007gg}
S.~Fay, R.~Tavakol and S.~Tsujikawa, Phys.\ Rev.\  D {\bf 75},
(2007) 063509.

\bibitem{Nojiri:2006be}
S.~Nojiri and S.~D.~Odintsov,
  Phys.\ Rev.\  D {\bf 74} (2006) 086005;
S.~Nojiri and S.~D.~Odintsov,
  J.\ Phys.\ Conf.\ Ser.\  {\bf 66}, 012005 (2007);
  S.~Capozziello, S.~Nojiri, S.~D.~Odintsov and A.~Troisi,
  Phys.\ Lett.\  B {\bf 639} (2006) 135;
S.~Nojiri and S.~D.~Odintsov, arXiv:0804.3519 [hep-th].

\bibitem{Cognola:2006eg}
  G.~Cognola, E.~Elizalde, S.~Nojiri, S.~D.~Odintsov and S.~Zerbini,
  Phys.\ Rev.\  D {\bf 73}, 084007 (2006)
  %[arXiv:hep-th/0601008].

\bibitem{darkmatter} S.~Capozziello, V.~F.~Cardone and A.~Troisi,
JCAP {\bf 0608}, 001 (2006); S. Capozziello, V. F. Cardone and A.
Troisi, Mon. Not. R. Astron. Soc. {\bf 375}, 1423 (2007);
A.~Borowiec, W.~Godlowski and M.~Szydlowski, Int. J. Geom. Meth.
Mod. Phys. {\bf 4} (2007) 183; C.~F.~Martins and P.~Salucci,
  Mon.\ Not.\ Roy.\ Astron.\ Soc.\  {\bf 381}, 1103 (2007);
%
  C.~G.~Boehmer, T.~Harko and F.~S.~N.~Lobo,
  Astropart. Phys. {\bf 29}, 386-392 (2008);
  %arXiv:0709.0046 [gr-qc];
  %%CITATION = ARXIV:0709.0046;%%
C.~G.~Boehmer, T.~Harko and F.~S.~N.~Lobo,
  JCAP {\bf 0803}, 024 (2008);
  %%CITATION = JCAPA,0803,024;%%
F.~S.~N.~Lobo,
  %``The dark side of gravity: Modified theories of gravity,''
  arXiv:0807.1640 [gr-qc].
  %%CITATION = ARXIV:0807.1640;%%


\bibitem{Nojiri:2004bi}
  S.~Nojiri and S.~D.~Odintsov,
  Phys.\ Lett.\  B {\bf 599} (2004) 137;
G.~Allemandi, A.~Borowiec, M.~Francaviglia and S.~D.~Odintsov,
 Phys.\ Rev.\  D {\bf 72} (2005) 063505;
  T.~Koivisto,
  Class.\ Quant.\ Grav.\  {\bf 23}, (2006) 4289;
  O.~Bertolami, C.~G.~B\"ohmer, T.~Harko and F.~S.~N.~Lobo,
  Phys. Rev. D {\bf 75} (2007) 104016;
O. Bertolami and J. P\'aramos, Phys.\ Rev.\  D {\bf 77}, 084018
(2008); V.~Faraoni, Phys. Rev. D {\bf 76}, 127501 (2007); O.
Bertolami and J. P\' aramos, arXiv:0805.1241 [gr-qc];
%
T.~P.~Sotiriou, arXiv:0805.1160 [gr-qc]; T.~P.~Sotiriou and
V.~Faraoni, arXiv:0805.1249 [gr-qc];
%
O. Bertolami, F. S. N. Lobo and J. P\'{a}ramos, Phys. Rev. D {\bf
78}, 064036 (2008).
  %``Nonminimal coupling of perfect fluids to curvature,''
  %arXiv:0806.4434 [gr-qc].

\bibitem{Olmo:2006zu}
  G.~J.~Olmo, Phys.\ Rev.\ Lett.\  {\bf 98}, (2007) 061101.

\bibitem{viablemodels}
  M.~Amarzguioui, O.~Elgaroy, D.~F.~Mota and T.~Multamaki,
 Astron.\ Astrophys.\  {\bf 454} (2006) 707;
  L.~Amendola, D.~Polarski and S.~Tsujikawa,
  Phys.\ Rev.\ Lett.\  {\bf 98}, (2007) 131302;
  L.~Amendola, R.~Gannouji, D.~Polarski and S.~Tsujikawa,
  Phys.\ Rev.\  D {\bf 75}, (2007) 083504;
  T.~Koivisto,
  Phys.\ Rev.\  D {\bf 76}, 043527 (2007);
  %%CITATION = PHRVA,D76,043527;%%
  A.~A.~Starobinsky,
  JETP Lett.\  {\bf 86}, 157 (2007).
  %%CITATION = JTPLA,86,157;%%

\bibitem{Hu:2007nk}
  W.~Hu and I.~Sawicki,
  Phys.\ Rev.\  D {\bf 76}, 064004 (2007).
  %%CITATION = PHRVA,D76,064004;%%

\bibitem{Sokolowski:2007pk}
  L.~M.~Sokolowski,
  Class. Quant. Grav. {\bf 24} 3391-3411 (2007).
  %[gr-qc/0702097].

\bibitem{solartests}
  T.~Chiba,
  Phys.\ Lett.\ B {\bf 575}, (2003) 1;
  A.~L.~Erickcek, T.~L.~Smith and M.~Kamionkowski,
  Phys.\ Rev.\  D {\bf 74}, (2006) 121501(R);
  T.~Chiba, T.~L.~Smith and A.~L.~Erickcek,
  Phys.\ Rev.\  D {\bf 75}, 124014 (2007);
  G.~J.~Olmo,  Phys.\ Rev.\  D {\bf 75} (2007) 023511.

\bibitem{solartests2}
  S.~Nojiri and S.~D.~Odintsov,
  Phys.\ Rev.\  D {\bf 68} (2003) 123512;
  V.~Faraoni, Phys.\ Rev.\  D {\bf 74} (2006) 023529;
  T.~Faulkner, M.~Tegmark, E.~F.~Bunn and Y.~Mao,
  Phys.\ Rev.\  D {\bf 76}, 063505 (2007).

\bibitem{Sawicki:2007tf}
  I.~Sawicki and W.~Hu,
  Phys.\ Rev.\  D {\bf 75}, 127502 (2007).

\bibitem{Amendola:2007nt}
  L.~Amendola and S.~Tsujikawa,
  Phys.\ Lett.\  B {\bf 660}, 125 (2008).

\bibitem{Faraoni:2006sy}
  V.~Faraoni,
  Phys.\ Rev.\  D {\bf 74}, (2006) 104017;
  S.~Carloni, P.~K.~S.~Dunsby, S.~Capozziello and A.~Troisi,
  Class.\ Quant.\ Grav.\  {\bf 22} (2005) 4839;
  J.~A.~Leach, S.~Carloni and P.~K.~S.~Dunsby,
  Class.\ Quant.\ Grav.\  {\bf 23} (2006) 4915;
  S.~Carloni, A.~Troisi and P.~K.~S.~Dunsby,
  arXiv:0706.0452 [gr-qc].

\bibitem{Boehmer:2007tr}
C.~G.~B\"{o}hmer, L.~Hollenstein and F.~S.~N.~Lobo,
  Phys.\ Rev.\  D {\bf 76}, 084005 (2007);
R.~Goswami, N.~Goheer and P.~K.~S.~Dunsby,
  arXiv:0804.3528 [gr-qc].

\bibitem{deSitter}
  G.~Cognola, E.~Elizalde, S.~Nojiri, S.~D.~Odintsov and S.~Zerbini,
  JCAP {\bf 0502} (2005) 010;
  V.~Faraoni, Phys.\ Rev.\  D {\bf 72}, (2005) 061501(R);
  V.~Faraoni, S.~Nadeau, Phys.\ Rev.\  D {\bf 72}, (2005) 124005;
  V.~Faraoni, Phys.\ Rev.\  D {\bf 75}, (2007) 067302;
  G.~Cognola, M.~Gastaldi and S.~Zerbini,
  Int.\ J.\ Theor.\ Phys.\  {\bf 47}, 898 (2008).

\bibitem{Nojiri:2006ri}
  S.~Nojiri and S.~D.~Odintsov,
  Int.\ J.\ Geom.\ Meth.\ Mod.\ Phys.\  {\bf 4} (2007) 115.

\bibitem{structureform}
  T.~Koivisto and H.~Kurki-Suonio,
  Class.\ Quant.\ Grav.\  {\bf 23}, (2006) 2355;
  R.~Bean, D.~Bernat, L.~Pogosian, A.~Silvestri and M.~Trodden,
  Phys.\ Rev.\  D {\bf 75}, (2007) 064020.

\bibitem{Tsuj}
  S.~Tsujikawa, Phys.\ Rev.\  D {\bf 76}, 023514 (2007).

\bibitem{Uddin:2007gj}
  K.~Uddin, J.~E.~Lidsey and R.~Tavakol,
  Class.\ Quant.\ Grav.\  {\bf 24}, 3951 (2007).

\bibitem{Bazeia:2007jj}
  D.~Bazeia, B.~Carneiro da Cunha, R.~Menezes and A.~Y.~Petrov,
  Phys.\ Lett.\  B {\bf 649} (2007) 445.

\bibitem{Clifton:2005aj}
  T.~Clifton and J.~D.~Barrow,
  %``The Power of General Relativity,''
  Phys.\ Rev.\  D {\bf 72}, 103005 (2005).
  %[arXiv:gr-qc/0509059].
  %%CITATION = PHRVA,D72,103005;%%

\bibitem{SSSsol}
  T.~Multamaki and I.~Vilja, Phys.\ Rev.\  D {\bf 76}, 064021 (2007);
  K.~Kainulainen, J.~Piilonen, V.~Reijonen and D.~Sunhede,
  Phys.\ Rev.\  D {\bf 76}, 024020 (2007);
  S.~Capozziello, A.~Stabile and A.~Troisi,
  Class.\ Quant.\ Grav.\  {\bf 24}, (2007) 2153;
  M.~D.~Seifert,
  Phys.\ Rev.\  D {\bf 76}, 064002 (2007).

\bibitem{Multamaki:2006zb}
T.~Multamaki and I.~Vilja, Phys.\ Rev.\ D {\bf 74}, 064022 (2006).

\bibitem{Multamaki:2006ym}
  T.~Multamaki and I.~Vilja,
  Phys.\ Rev.\  D {\bf 76}, 064021 (2007).

\bibitem{Capozziello:2007wc}
  S.~Capozziello, A.~Stabile and A.~Troisi,
  Class.\ Quant.\ Grav.\  {\bf 24}, 2153 (2007).

\bibitem{Capozziello:2007ms}
  S.~Capozziello, A.~Stabile and A.~Troisi,
  Phys.\ Rev.\  D {\bf 76}, 104019 (2007).

\bibitem{Kainulainen:2007bt}
  K.~Kainulainen, J.~Piilonen, V.~Reijonen and D.~Sunhede,
  Phys.\ Rev.\  D {\bf 76}, 024020 (2007).

\bibitem{Henttunen:2007bz}
  K.~Henttunen, T.~Multamaki and I.~Vilja,
  Phys.\ Rev.\  D {\bf 77}, 024040 (2008).

\bibitem{Multamaki:2007jk}
  T.~Multamaki and I.~Vilja,
  Phys.\ Lett.\  B {\bf 659}, 843 (2008).

\bibitem{Dymnikova2}
I.~Dymnikova, Class.\ Quant.\ Grav.\  {\bf 21}, 4417 (2004).

\bibitem{MannKaz}
P. D. Mannheim and D. Kazanas, Astroph. Journ. {\bf 342}, 635
(1989); D. Kazanas and P. D. Mannheim, Astroph. Journ. Supp.
Series {\bf 76}, 421 (1991).

\bibitem{Bamba:2008ja}
 K.~Bamba and S.~D.~Odintsov, JCAP {\bf 0804}, 024 (2008).

\bibitem{BI}
M. Born, Proc. Roy. Soc. Lond. {\bf A143}, 410 (1934); M. Born and
L. Infeld, Proc. Roy. Soc. {\bf A144}, 425 (1934).

\bibitem{Pleb}
J. F. Pleba\'{n}ski, ``Lectures on non-linear electrodynamics,''
monograph of the Niels Bohr Institute Nordita, Copenhagen (1968).

\bibitem{Witten}
N.~Seiberg and E.~Witten, JHEP {\bf 9909}, 032 (1999).

\bibitem{cosmoNLED}
R.~Garcia-Salcedo and N.~Breton, Int.\ J.\ Mod.\ Phys.\  A {\bf
15}, 4341 (2000); R.~Garcia-Salcedo and N.~Breton, Class.\ Quant.\
Grav.\  {\bf 20}, 5425 (2003); R.~Garcia-Salcedo and N.~Breton,
Class.\ Quant.\ Grav.\  {\bf 22}, 4783 (2005); V. V. Dyadichev, D.
V. Gal'tsov, A. G. Zorin and M. Yu. Zotov, Phys. Rev. D {\bf 65},
084007 (2002); D. N. Vollick, Gen. Rel. Grav. {\bf 35}, 1511-1516
(2003).

\bibitem{Novello}
M.~Novello, S.~E.~Perez Bergliaffa and J.~Salim,
Phys.\ Rev.\  D {\bf 69}, 127301 (2004).


\bibitem{Garcia}
E. Ay\'{o}n-Beato and A. Garc\'{i}a, Phys. Rev. Lett. {\bf 80},
5056-5059 (1998).

\bibitem{Garcia2}
E. Ay\'{o}n-Beato and A. Garc\'{i}a, Phys. Lett. B {\bf 464}, 25
(1999); E. Ay\'{o}n-Beato and A. Garc\'{i}a, Gen. Rel. Grav. {\bf
31}, 629-633 (1999).

\bibitem{Bronnikov1}
K. A. Bronnikov, Phys. Rev. D {\bf 63}, 044005 (2001).

\bibitem{Dymnikova}
I. Dymnikova, Class. Quant. Grav. {\bf 21}, 4417-4429 (2004).

\bibitem{Arellano1}
A.~V.~B.~Arellano and F.~S.~N.~Lobo, Class.\ Quant.\ Grav.\  {\bf
23}, 5811 (2006);
A.~V.~B.~Arellano and F.~S.~N.~Lobo,
Class.\ Quant.\ Grav.\  {\bf 23}, 7229 (2006);
A.~V.~B.~Arellano, N.~Breton and R.~Garcia-Salcedo,
arXiv:0804.3944 [gr-qc].

\bibitem{Arellano3}
F.~S.~N.~Lobo and A.~V.~B.~Arellano,
Class.\ Quant.\ Grav.\  {\bf 24}, 1069 (2007).

\bibitem{Jacobson:2007tj}
  T.~Jacobson,
  %``When is g_{tt} g_{rr} = -1?,''
  Class.\ Quant.\ Grav.\  {\bf 24}, 5717 (2007).
  %[arXiv:0707.3222 [gr-qc]].
  %%CITATION = CQGRD,24,5717;%%



\end{thebibliography}
\end{document}